\documentclass[a4paper,fleqn,usenatbib,useAMS]{mpazh}
\pdfoutput=1

\usepackage{hyperref}

\usepackage[utf8]{inputenc}
\usepackage{graphicx}
\usepackage{grffile}
\usepackage{natbib}

\usepackage{floatrow}
\usepackage{caption}
\usepackage{longtable}
\usepackage{amssymb}

\usepackage{multirow}

\usepackage[utf8]{inputenc}
\usepackage{graphicx}

\usepackage{floatrow}
\usepackage{longtable}
\usepackage{booktabs}

\usepackage[toc]{appendix}

\usepackage{multirow}

\usepackage[toc]{appendix}

\usepackage{array}

\def\ps1{\emph{Pan-STARRS1}}

\def\srg{SRG}
\def\art{ART-XC}
\def\ero{eROSITA}
\def\srge{SRG/eROSITA}
\def\xmm{XMM-Newton}

\def\einstein{Einstein}

\def\lx{L_{\rm X}}
\def\mbh{M_{\rm BH}}

\def\fmean{\langle f\rangle}

\def\pg{PG\,1634+706}

\title{
    X-ray Properties of the Luminous Quasar \pg\ at $z=1.337$ from SRG and XMM-Newton Data
}

\author{
G.~S.~Uskov\address{1}\email{uskov@cosmos.ru},
S.~Yu.~Sazonov\address{1},
M.~R.~Gilfanov\address{1,2},
I.~Yu.~Lapshov\address{1},
R.~A.~Sunyaev\address{1,2}
  \addresstext{1}{\it Space Research Institute, Russian Academy of Sciences, Moscow, 117997 Russia}
  \addresstext{2}{\it Max Planck Institut fur Astrophysik, Karl-Schwarzschild-Str. 1, Postfach 1317, D-85741 Garching, Germany}
}

\author[Uskov et. al.]{
	G.S. Uskov,
			\thanks{\href{mailto:uskov@cosmos.ru}{\nolinkurl{uskov@cosmos.ru}}}
				$^{1}$
		S. Yu. Sazonov,
				$^{1}$
        M. R. Gilfanov,
				$^{1,2}$
    \newauthor
        I. Yu. Lapshov,
                $^{1}$
        R. A. Sunyaev,
				$^{1,2}$
	\\
			$^{1}$Space Research Institute, Russian Academy of Sciences, Moscow,
117997 Russia\\
			$^{2}$Max Planck Institut for Astrophysik, Karl-Schwarzschild-Str. 1,
85741 Garching, Germany\\
}

\date{Accepted XXX. Received YYY; in original form ZZZ}

\pubyear{2023}

\begin{document}
\label{firstpage}
\pagerange{\pageref{firstpage}--\pageref{lastpage}}
\maketitle

\begin{abstract}
In the fall of 2019, during the in-flight calibration phase of the SRG observatory, the onboard \ero\ and Mikhail Pavlinsky \art\ telescopes carried out a series of observations of PG\,1634+706 — one of the most luminous (an X-ray luminosity $\sim 10^{46}$~erg/s) quasars in the Universe at $z<2$. 
Approximately at the same dates this quasar was also observed by the \xmm\ observatory.
Although the object had already been repeatedly studied in X-rays previously, its new observations allowed its energy spectrum to be measured more accurately in the wide range 1--30~keV (in the quasar rest frame).
Its spectrum can be described by a two-component model that consists of a power-law continuum with a slope $\Gamma\approx 1.9$ and a broadened iron emission line at an energy of about 6.4~keV.
The X-ray variability of the quasar was also investigated.
On time scales of the order of several hours (here and below, in the source rest frame) the X-ray luminosity does not exhibit a statistically significant variability.
However, it changed noticeably from observation to observation in the fall of 2019, having increased approximately by a factor of 1.5 in 25 days.
A comparison of the new \srg\ and \xmm\ measurements with the previous measurements of other X-ray observatories has shown that in the entire 17-year history of observations of the quasar \pg\ its X-ray luminosity has varied by no more than a factor of 2.5, while the variations on time scales of several weeks and several years are comparable in amplitude.


\end{abstract}

\section{INTRODUCTION}
\label{s:intro}

Active galactic nuclei (AGNs) manifest themselves in a wide wavelength range, from radio to gamma-rays.
A significant fraction ($\sim 10$\%, see, e.g., \citealt{elvis1994,sazonov2004,vasudevan2007,shang2011,sazonov2012}) of the bolometric luminosity of AGNs is accounted for by the X-ray emission.
This emission is believed to originate in the hot corona of the accretion disk as a result of the Comptonization of thermal (mainly ultraviolet) radiation from the disk.
In addition, features associated with the reflection of hard coronal radiation from the disk and the surrounding dusty torus are observed in the X-ray spectra of AGNs (for a recent review, see, e.g., \citealt{malizia2020}).
Thus, the X-rays from AGNs carry important information about the accretion of matter onto supermassive black holes (SMBHs).

Based on data from space observatories, the X-ray spectra of many Seyfert galaxies, i.e., AGNs with a comparatively low luminosity ($\lx\lesssim 10^{44}$\,erg/s in the 2--10\,keV band) located in the nearby Universe ($z\lesssim 0.1$), have been studied in detail (see, e.g., \citealt{derosa2012,ricci2017}).
As a rule, such objects are characterized by comparatively low black hole masses ($\mbh\lesssim 10^9$\,$M_\odot$) and accretion rates (less than 10\% of the critical one at which the Eddington luminosity is reached; see, e.g., \citealt{khorunzhev2012,prokhorenko2021,ananna2022}).
Since the properties of the accretion disks and their coronas can depend strongly on both black hole mass and accretion rate \citep{shakura1973}, it is important to investigate the X-ray emission of not only Seyfert galaxies but also AGNs with more massive black holes and/or higher Eddington ratios.
In particular, of considerable interest are narrow-line Seyfert 1 galaxies, which are characterized by higher (close to the critical one) accretion rates than those for ordinary Seyfert galaxies at comparatively low black hole masses.
Such objects have been well studied, and one of their distinctive features is comparative softness of their X-ray spectra (see, e.g., \citealt{jin2012}).

\begin{figure}
    \centering
    \includegraphics[width=1\columnwidth]{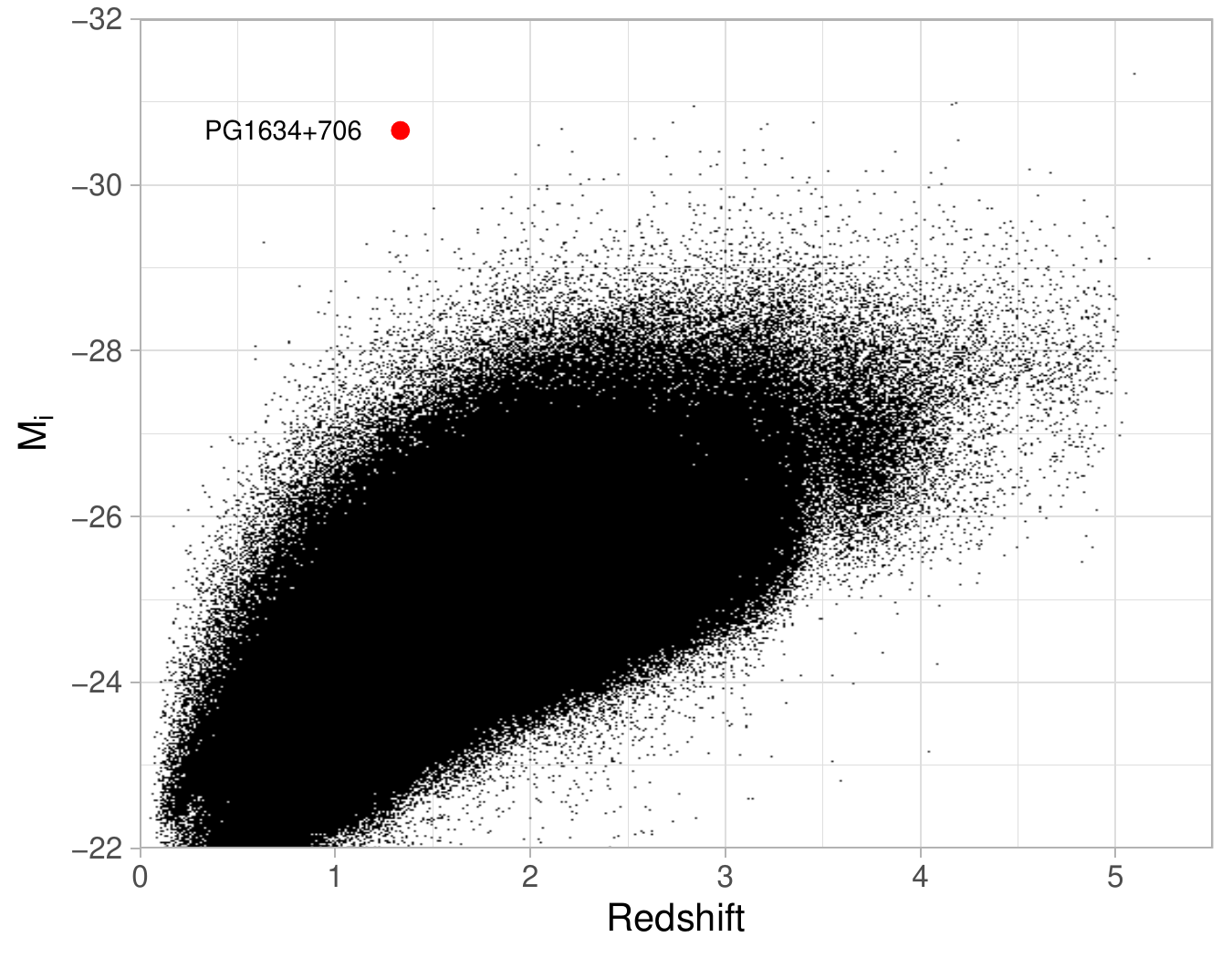}
    \caption{
    Position of the quasar \pg\ (red dot) on the redshift \emph{i}$[z=2]$ absolute magnitude diagram.
    The black dots indicate the objects from the 16th data release of the SDSS quasar catalog (DR16v4, \citealt{lyke2020}).
    The absolute magnitude of the quasar \pg, $M_i=-30.65$, was estimated using Eq. (4) 
    from \citealt{richards2006} based on the luminosity measurement at a wavelength of 2500\AA 
    given in Table 2 from \citealt{shemmer2014}.
    It should be noted that the quasar \pg\ does not enter into the DR16v4 catalog, since this sky region was not covered during the spectroscopic SDSS.
    }
    \label{fig:sdss}
\end{figure}

Of special interest are high-luminosity ($\lx\gtrsim10^{46}$\,erg/s) quasars in which a regime of accretion onto a black hole with a mass $\mbh\gtrsim 10^9$\,$M_\odot$ close to the Eddington limit is probably realized.
Such objects are very rare in the Universe, and even the nearest of them are at enormous distances from us, which makes their X-ray spectroscopy very difficult.
Therefore, one of the most interesting quasars is \pg. This quasar is characterized by huge X-ray ($\lx\sim 10^{46}$\,erg/s) and bolometric ($\sim 10^{48}$\,erg/s) luminosities (see, e.g., \citealt{shemmer2014}\footnote{In the cited paper a luminosity estimate at a wavelength of 2500\AA, $\nu L_{\nu,2500} \sim 3\times 10^{47}$~erg/s, is given, while the corresponding bolometric correction for quasars is estimated to be $\sim 3$ \citep{krawczyk2013}.}), but, at the same time, it is located at a moderate redshift, $z=1.337$ (\citealt{neeleman2016}, see Fig.~\ref{fig:sdss}).
The black hole mass is estimated to be $\sim 10^{10}$\,$M_\odot$ (a rough estimate based on the H$\beta$ width and luminosity), while the accretion rate is close to the critical one \citep{kelly2008}.
The quasar was discovered by the ultraviolet excess during the famous Palomar–Green survey \citep{green1986}, and its first studies in X-rays with the \einstein\ orbital observatory were carried out already in the early 1980s \citep{tananbaum1986}.
Since then it has repeatedly become a target of X-ray observations, which has allowed its spectral characteristics and variability to be investigated.

The Russian Spectrum–RG X-ray observatory (hereafter \srg, \citealt{sunyaev2021}) was launched on July 13, 2019, from the Baikonur Cosmodrome.
During the flight of the spacecraft to the Lagrange point L2 of the Sun--Earth system in July--December 2019 the onboard \ero\ \citep{predehl2021} and Mikhail Pavlinsky \art\ \citep{pavlinsky2021} telescopes observed a number of astrophysical objects to verify the performance of the instruments and to calibrate them.
The quasar \pg, which was observed several times in the fall of 2019, was also among the targets.
Its \xmm\ observations \citep{jansen2001} were also carried approximately at the same dates.
The data obtained from the three telescopes allowed us to investigate the X-ray spec- trum of \pg\ and to study its variability with high reliability and accuracy.
The results of this investigation are presented in our paper.

We use a $\Lambda$CDM cosmological model with parameters $H_0=70$~km/s/Mpc and $\Omega_\Lambda = 0.7$.

\begin{table*}
\renewcommand{\tabcolsep}{0.1cm}
\caption{\label{tab:obs_2019} 
Log of observations of \pg\ in 2019
}
\centering
\begin{tabular}[t]{llll}
\toprule
Period of observations & Observatory & Telescopes, instruments & Exposure time, ks\\
\midrule
2019-09-29 21:38 -- 09-30 13:05 & \srg\ & \art, \ero\ (TM6) & 48.3, 55.0  \\
2019-10-20 14:42 -- 10-21 01:52 & \srg\ & \art, \ero\ (TM1-7) & 37.7, 39.0 \\
2019-10-25 18:02 -- 10-26 00:58 & \xmm\ & EPIC-PN & 14.4\\
2019-11-23 09:11 -- 20:18 & \srg\ & \art, \ero\ (TM1-7) & 37.0, 38.3\\
2019-11-24 14:38 -- 19:54 & \xmm\ & EPIC-PN & 12.7\\
2019-11-26 07:05 -- 18:14 & \srg\ & \ero (TM1-7) & 39.8\\
2019-11-26 14:32 -- 22:19 & \xmm\ & EPIC-PN & 17.3\\
\bottomrule
\end{tabular}
\begin{flushleft}
The first column gives the universal times of the start and end of the observation.
The second and third columns give the name of the observatory and the corresponding telescopes and instrument; the switched-on cameras for \ero\ are given in parentheses.
The last column gives the effective exposure time corrected for vignetting.
\end{flushleft}
\end{table*}

\section{OBSERVATIONS AND DATA ANALYSIS}
\label{s:data}

Table~\ref{tab:obs_2019} presents information about the X-ray observations of the quasar \pg\ in the fall of 2019 whose data were used in this paper.
These include four \srg\ observations, in which \pg\ was no farther than 13\arcmin\ from the \ero\ axis, and three \xmm\ observations.
The latter were approximately synchronized with the \srg\ observations: the difference in the epochs of observations is about 5 days for the October observation and less than one day for the two November observations.
During the \srg\ observation on September 29 only the sixth (TM6) of the seven \ero\ modules was switched on, while in the three later observations already all seven telescope cameras (TM1--TM7) operated.

Since the \srg\ observations under consideration were carried out at the Calibration and Performance Verification (Cal-PV) phase, their target, the quasar \pg\, was at different angular distances from the optical axis of the \ero\ and \art\ telescopes (approximately aligned) in different observations.
More specifically, the source was almost at the center of the field of view in the first observation, at 2--3 arcmin from the axis in the second and third ones, and approximately at 13 arcmin from the center in the last one.
Given the significant drop in the effective area of the \art\ telescope at large angular distances from the axis, we used the data from the latest \ero\ observation (November 26) only to investigate the variability of the X-ray flux and did not use the \ero\ and \art\ data from this observation for our spectral analysis.

For all three \xmm\ observations we used only the data from the EPIC-PN camera (hereafter EPN), the most sensitive instrument of the observatory.
No data from the EPIC-MOS cameras were used, since the observations of \pg\ in the fall of 2019 were carried out in the mode of a small window with an angular size of $258\arcsec\times 258\arcsec$ for EPN and $110\arcsec\times 110\arcsec$ for EPIC-MOS, which makes a reliable determination of the background (for our spectral analysis) in the latter case virtually impossible.

\subsection{\srge}

\begin{figure*}
    \centering
    \includegraphics[width=.32\columnwidth]{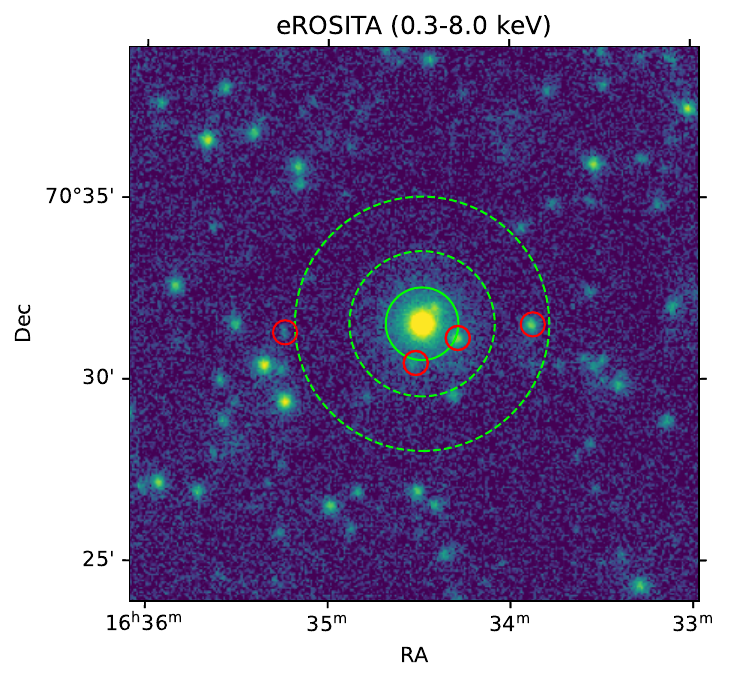}
    \includegraphics[width=.32\columnwidth]{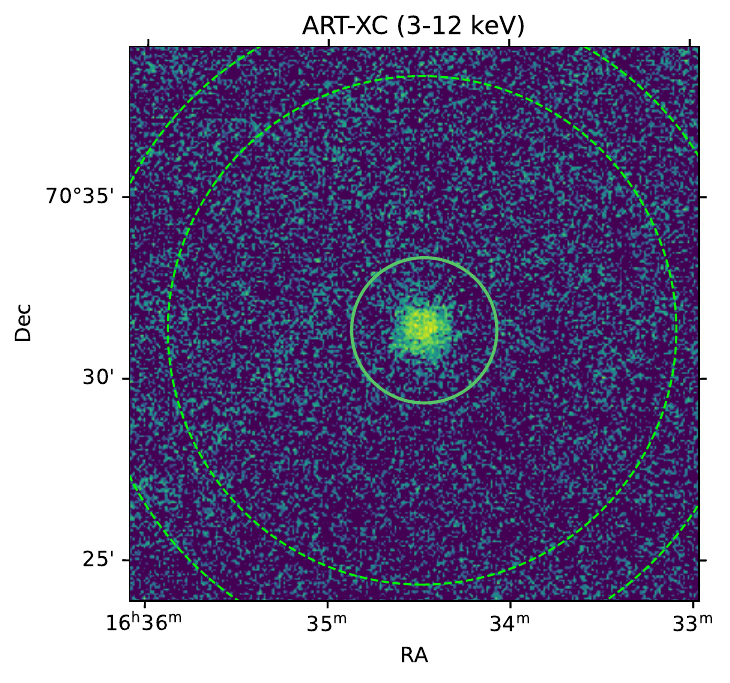}
    \includegraphics[width=.32\columnwidth]{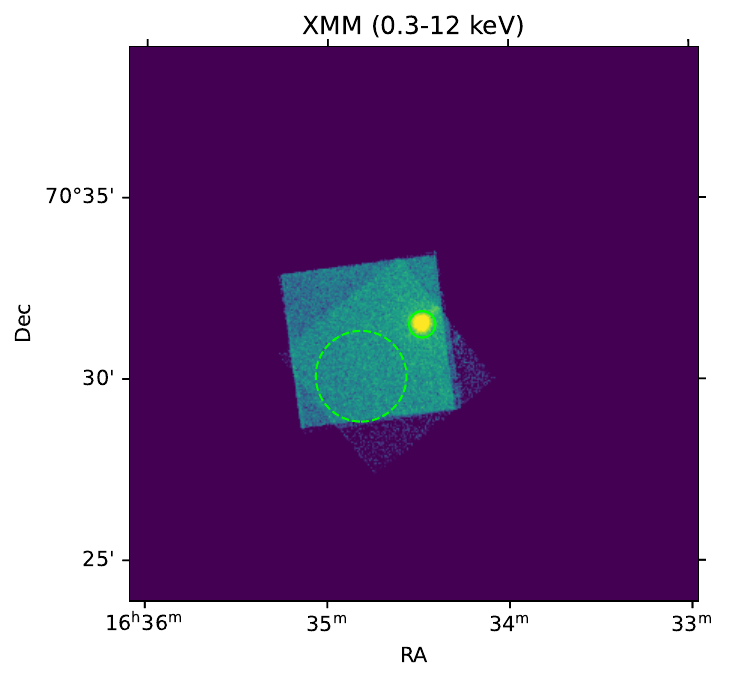}
    \caption{
    The X-ray images of the quasar \pg\ obtained from the series of observations (see Table~\ref{tab:obs_2019}) with the \ero\ and \art\ telescopes of the \srg\ observatory and the EPN telescope of the \xmm\ observatory (in the 0.3--8, 3--12, and 0.3--12 keV energy bands, respectively).
    On each panel the green solid circumference marks the source spectrum extraction region, while the dashed circumferences mark the background extraction region (for \xmm\ these regions slightly differ in different observations; the regions for the observation on October 25, 2019, OBSID=0852980501, are shown).
    The red circles in the \ero\ image indicate the regions around other detected sources the events in which were excluded from our spectral and timing analyses.
    }
    \label{fig:img_ero}
\end{figure*}

We calibrated and processed the \ero\ data using the eROSITA Science Analysis Software System (\textit{eSASS}) software package of version esass\_211201 and the software developed by us. The calibration database of version caldb\_211201 was used.

The lists of events were filtered using the QUALGTI good time intervals with the \textit{evtool} (v2.29.2/2.18)\footnote{erosita.mpe.mpg.de/edr/DataAnalysis/evtool\_doc} code.
We excluded the time intervals during which the \srg\ observatory was slewed.
The pure vignetting-corrected exposure time for each observation is given in Table~\ref{tab:obs_2019}.

The \textit{srctool} procedure was used to extract the X-ray spectra and light curves. The photons from the source were extracted from a region with a radius of 1\arcmin\ (Fig.~\ref{fig:img_ero}). The background was estimated in a ring around the source with inner and outer radii of 2\arcmin\ and 3.5\arcmin\, respectively. The events within 20\arcsec\ of other sources closest to the quasar detected by \ero\ (Fig.~\ref{fig:img_ero}) were excluded from consideration.

The spectra of the source obtained were binned in such a way that there were at least 30 counts (as for \art, see below) from the source and the background in each energy bin. The light curves were binned in 1~ks, except for the September 29, 2019 observation that was binned in 4~ks. With such binning there are at least 15 total counts from the source and the background in each time bin. At the end of each observation we removed the bins related to "bad" time (according to GTI). To calculate the 68\% confidence intervals for the count rate, we used the Monte Carlo method whereby the counts from the source and the background are randomly selected according to the Poisson distribution.

\subsection{\srg/\art}

The spectra were extracted with the standard \art\ software package adapted to the goals of this study using the current version of the calibrations. In individual energy intervals we constructed the photon, exposure, and vignetting maps. To take into account the contribution of the particle background to the source spectrum, we constructed the model particle background maps obtained during the all-sky survey in periods when no sources were observed in the field of view. The spectrum of the quasar \pg\ was extracted in a circle of radius 2\arcmin. The background normalization was estimated in a ring with inner and outer radii of 7\arcmin\ and 9\arcmin\, respectively (Fig.~\ref{fig:img_ero}). In the ring region the model background map was normalized to the background map with allowance made for the exposure time. The normalized background map was used in the region with the source to subtract the underlying background. The remaining photons were assumed to belong to the source and were corrected for the vignetting map for the reduction to the nominal effective area.

The angular resolution of \art\ does not allow one to reliably separate the emissions from the quasar \pg\ and other point sources that are detected by \ero\ within 2\arcmin\ of it (Fig.~\ref{fig:img_ero}). However, according to the \ero\ data, the contribution of these sources to the flux from the quasar in the energy range from 0.3 to 8~keV does not exceed 5\%. Consequently, the unresolved sources should not affect noticeably the spectrum measured by the \art\ telescope.
We restricted ourselves to using the energy channels above 5~keV when performing our spectral analysis. This is because the \art\ response matrix is not well calibrated near the lower boundary of the sensitivity range of its detectors ($\sim 3$--4~keV). The remaining energy channels were binned in such a way that there were at least 30 counts in each of the bins (this is dictated by the necessity of using the $\chi^2$ test in modeling the \srg\ spectra, since the background has already been subtracted in the \art\ spectral data data used by us).

\subsection{\xmm/EPN}

The primary processing of the data from the EPN camera of the \xmm\ observatory was performed using the Science Analysis System (SAS) v20.0 software. We used the latest calibration versions at the SAS v20.0 release time.

The file of events was filtered: only the single and double events (PATTERN $\leq$ 4) in the energy range 0.3--12~keV were left; the times with an enhanced background\footnote{www.cosmos.esa.int/web/xmm-newton/sas-thread-epic-filterbackground} and the events in and near bad pixels (FLAG=0) were removed.

The spectral and timing data were extracted with the xmmselect code. The source and background regions were chosen so as to maximize the signal-to-noise ratio. The right panel in Fig.~\ref{fig:img_ero} shows the combined image from the three 2019 \xmm\ observations.

The spectra of the source were binned in such a way that there were at least five counts in each energy bin. This is needed to fit the spectra using the $W$-statistic (below in the text called Cstat), which takes into account the presence of an X-ray background with a Poisson distribution. The light curves were binned in 1~ks. We also removed the time bins with a fractional exposure FRACEXP < 0.1. With such binning there are at least 15 net counts from the source in each time bin.

\section{RAPID VARIABILITY}
\label{s:lcurve}

\begin{figure*}
    \centering
    \includegraphics[width=1\columnwidth]{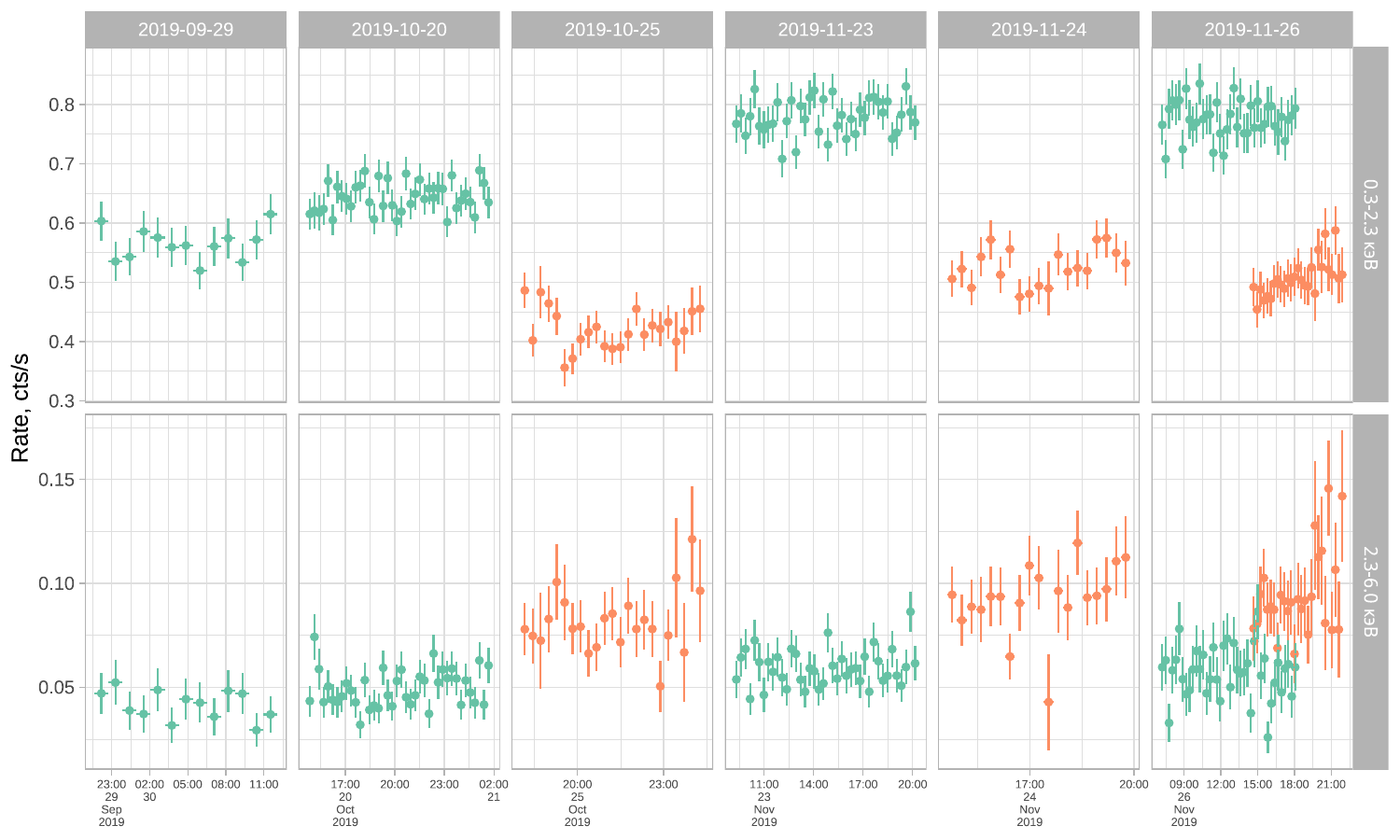}
    \caption{
    The \srg/\ero\ (green) and \xmm/EPN (orange) light curves (in counts per second) for the individual observations in two energy bands: 0.3--2.3 (top) and 2.3--6.0 keV (bottom). It should be noted that the absolute values of the \ero\ and EPN count rates cannot be directly compared with each other.
    }
    \label{fig:lc_ero}
\end{figure*}

\begin{table*}
\renewcommand{\arraystretch}{1.1}
\caption{\label{tab:time_stat} Characteristics of the rapid X-ray variability}
\centering
\begin{tabular}[t]{llllc}
\toprule
Telescope & Date & $\chi^2$ (dof) & $1-p$ & $\sigma_{\rm rms}^2$, $10^{-4}$\\
\midrule
\multicolumn{5}{c}{Energy range 0.3--2.3~keV}\\
\midrule
& 2019-09-29 & 8.3(12) & 0.759 & $-11\pm7$\\
 & 2019-10-20 & 33.9(39) & 0.702 & $-2\pm2$\\
 & 2019-11-23 & 38.7(39) & 0.482 & $-0.2\pm3$\\
\multirow[t]{-4}{*}{\raggedright\arraybackslash \ero} & 2019-11-26 & 31.9(39) & 0.782 & $-4\pm3$ \\
 & 2019-10-25 & 25.1(22) & 0.291 & $6\pm16$\\
 & 2019-11-24 & 17.4(18) & 0.499 & $-3\pm8$\\
\multirow[t]{-3}{*}{\raggedright\arraybackslash EPN} & 2019-11-26 & 18.1(26) & 0.872 & $-11\pm12$\\
\midrule
\multicolumn{5}{c}{Energy range 2.3--6.0~keV}\\
\midrule
 & 2019-09-29 & 7.1(12) & 0.853 & $-222\pm81$\\
 & 2019-10-20 & 45.6(39) & 0.217 & $63\pm67$\\
 & 2019-11-23 & 39(39) & 0.468 & $11\pm55$\\
\multirow[t]{-4}{*}{\raggedright\arraybackslash \ero} & 2019-11-26 & 52.7(39) & 0.070 & $60\pm111$\\
 & 2019-10-25 & 15(22) & 0.862 & $-108\pm104$\\
 & 2019-11-24 & 18.9(18) & 0.399 & $60\pm141$\\
\multirow[t]{-3}{*}{\raggedright\arraybackslash EPN} & 2019-11-26 & 22.9(26) & 0.637 & $56\pm118$ \\
\bottomrule
\end{tabular}
\end{table*}

To investigate the X-ray variability of the quasar \pg\ on time scales shorter than one day, we constructed the light curves from the \srg/\ero\ and \xmm/EPN data for the individual observations. Figure~\ref{fig:lc_ero} shows the time dependences of the count rate in two energy bands, 0.3--2.3 and 2.3--6.0~keV.

To understand whether there is a statistically significant variability of the X-ray flux in a given observation and a given energy band, we calculated the $\chi^2$ statistic:
\begin{equation}
    \chi^2 = \sum_{i=1}^{N} \frac{(f_{\rm i} - \overline{f})^2}{\sigma_{\rm i}^2},
\end{equation}
where $f_{\rm i}$ is the count rate in the {\it i}th time bin, $\sigma_{\rm i}$ is the corresponding error, $\overline{f}$ is the weighted mean count rate for the entire observation, and $N$ is the number of measurements. The probability to obtain by chance a value of the $\chi^2$ distribution greater than the measured one for ${\rm dof}=N-1$ degrees of freedom characterizes the probability ($1-p$) that the count rate was constant in a given observation.

To estimate the X-ray variability amplitude, we calculated the variance normalized to the mean flux and corrected for the flux measurement errors \citep{vaughan2003}:

\begin{equation}
    \sigma^2_{\rm rms} = \frac{S^2 - \overline{\sigma_{\rm err}^2}}{\fmean^2},
    \label{eq:sigma}
\end{equation}

where

\begin{equation}
    S^2 = \frac{1}{N-1} \sum_{i=1}^N (f_{\rm i} - \fmean)^2,
\end{equation}
\begin{equation}
    \overline{\sigma_{\rm err}^2} = \frac{1}{N}\sum_{\rm i}^N \sigma^2_{\rm i}
\end{equation}
and $\fmean$ is the arithmetic mean count rate.

The error in $\sigma_{\rm rms}^2$ can be estimated using the formula from \cite{turner1999}:
\begin{equation}
\delta\sigma_{\rm rms}^2=\frac{s_D}{\fmean^2\sqrt{N}},
\end{equation}

where

\begin{equation}
s^2_D = \frac{1}{N-1} \sum_{i=1}^{N}\{[(f_{\rm i} - \fmean)^2 - \sigma^2_{\rm i}] - \sigma^2_{\rm rms}\fmean^2\}^2.
\label{eq:sigmaerr}
\end{equation}

The derived values of $\chi^2$  (dof), $1-p$, and $\sigma_{\rm rms}^2$ are given in Table~\ref{tab:time_stat}. No statistically significant variability (at a confidence level greater than 2$\sigma$) was detected in any of the \srg/\ero\ and \xmm\ observations.

The light curves were also studied for the presence of a linear trend. Only for the \xmm\ observation on November 26, 2019, in the 0.3--2.3~keV energy band did we find a statistically significant ($p<0.0001$) improvement in the quality of the light-curve fit. More specifically, there is evidence for an increase in the count rate during the observation. At the same time, we did not find a similar positive trend in the \srg/\ero observation that was carried out only several hours earlier and partially overlaps in time with the \xmm\ observation.


\section{X-RAY SPECTRUM}
\label{s:spec}

\begin{figure*}
    \centering
    \includegraphics[width=0.9\columnwidth]{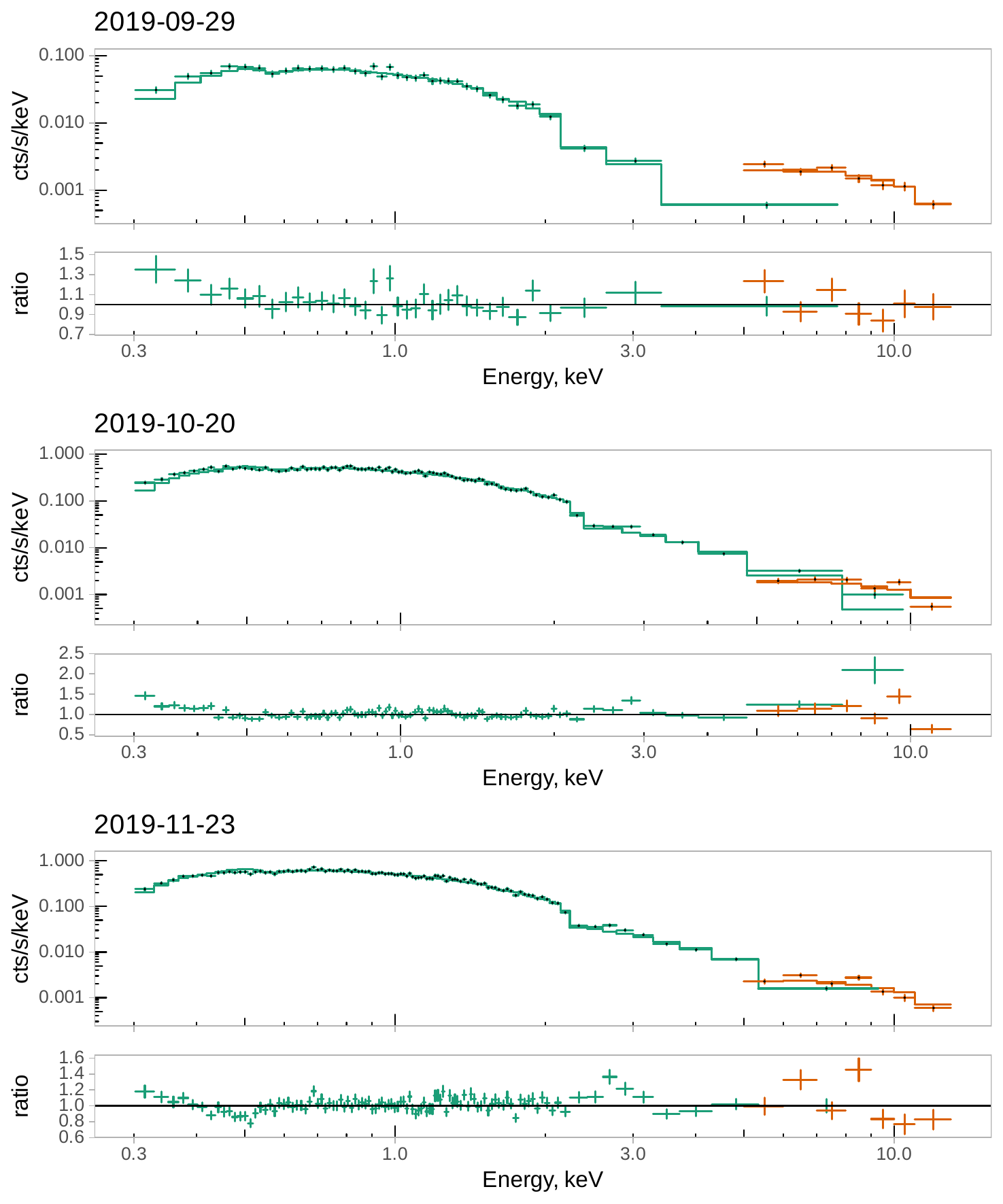}  
    \caption{ 
The X-ray spectrum of the quasar \pg\ measured with the \ero\ (green color) and \art\ (orange color) telescopes of the \srg\ observatory at different dates. All three spectra were fitted jointly by a power law with Galactic absorption with a single slope but different normalizations (see Table~\ref{tab:fit_pars_art_ero_xmm}). For better clarity, the spectra in the figure were rebinned. The data-to-model ratio is shown under each spectrum.
}
    \label{fig:art_ero_sum_spec}
\end{figure*}

\begin{figure*}
    \centering
   \includegraphics[width=0.9\columnwidth]{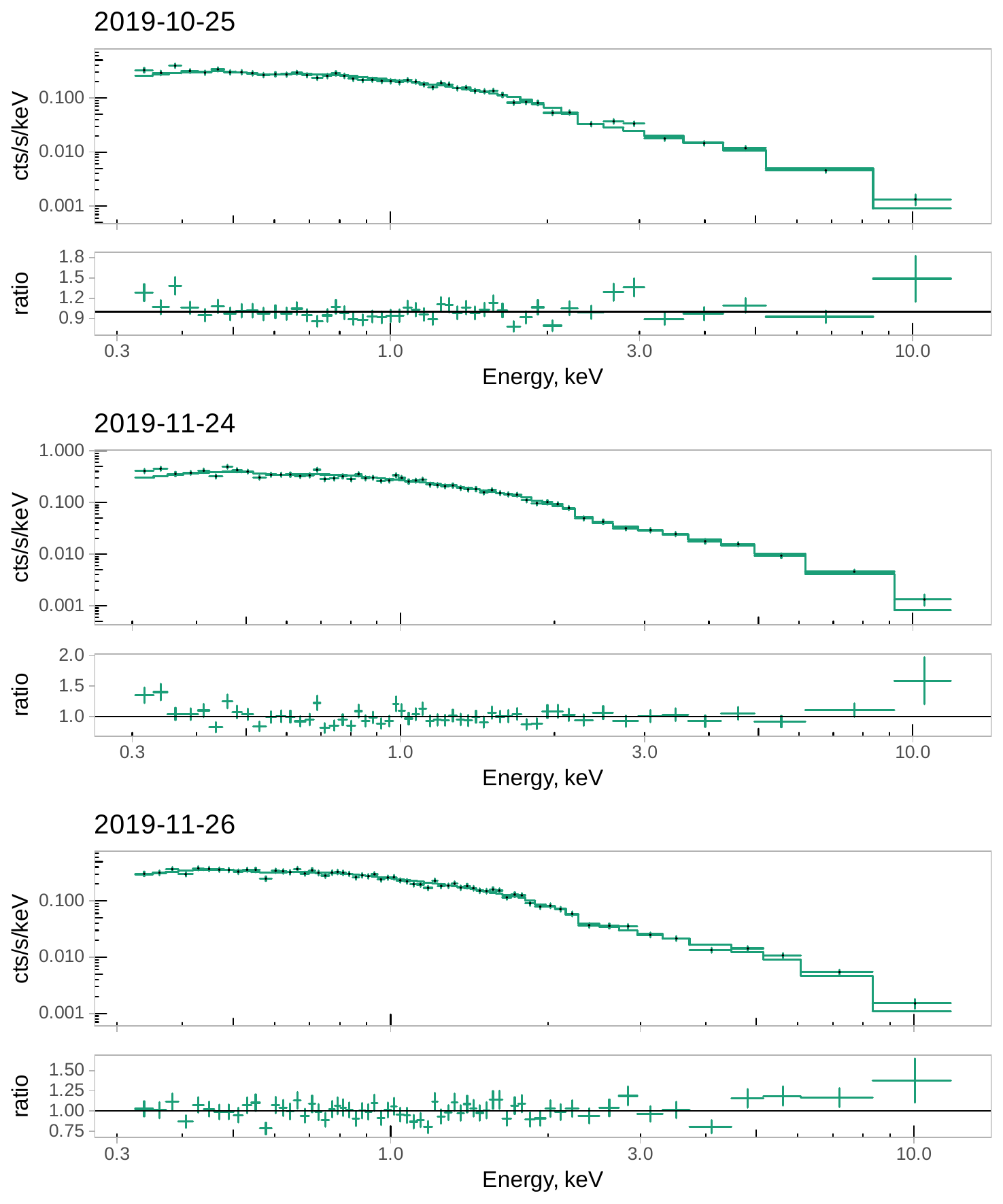}
    \caption{
    Same as Fig.~\ref{fig:art_ero_sum_spec} but for the \xmm\ data.
}
    \label{fig:xmm_sum_spec}
\end{figure*}

\begin{table*}
\caption{
\label{tab:fit_pars_art_ero_xmm} 
Results of fitting the measured \srg\ and \xmm\ spectra by a power law with absorption
}
\centering
\begin{tabular}[t]{llll}
\toprule
\multicolumn{4}{c}{\srg/\ero, \art} \\
\midrule
Date & 2019-09-29 & 2019-10-20 & 2019-11-23\\
Energy range, keV &  0.3--13 & 0.3--12 & 0.3--13\\
 \midrule
Counts & 4409.0+556.0 & 25108.8+395.0 & 29712.6+507.0\\
$\Gamma$ & \multicolumn{3}{c}{$1.843\pm0.008$} \\
$F_X$, $10^{-13}$~erg/s/cm$^2$ & $5.17_{-0.11}^{+0.11}$ & $5.92_{-0.09}^{+0.09}$ & $7.03_{-0.11}^{+0.11}$\\
Constant & $1.21\pm0.06$ & $0.96\pm0.05$ & $1.02\pm0.05$\\
$\chi^2$ (dof) & \multicolumn{3}{c}{1032 (790)} \\

\midrule
\multicolumn{4}{c}{XMM-Newton/EPN} \\
\midrule
Date & 2019-10-25 & 2019-11-24 & 2019-11-26\\
Energy range, keV & 0.3--11.8 & 0.3--11.8 & 0.3--11.8 \\
\midrule
Counts & 5872.8 & 6602.6 & 8269.2\\
$\Gamma$ & \multicolumn{3}{c}{$1.828\pm0.011$} \\
$F$ (4--12 keV), $10^{-13}$~erg/s/cm$^2$ & $6.42_{-0.16}^{+0.16}$ & $8.01_{-0.19}^{+0.20}$ & $7.72_{-0.19}^{+0.19}$\\
Cstat (dof) & \multicolumn{3}{c}{2101 (2133)} \\
\bottomrule
\end{tabular}
\begin{flushleft}
 Counts is the number of counts from the source minus the background (the sums of the \ero\ and \art\ counts are given for the \srg\ observations), $\Gamma$ is the spectral slope, $F$ is the flux in the observed 4--12~keV energy band corrected for Galactic absorption, Constant is the correction factor for the \art\ data relative to \ero, and dof is the number of degrees of freedom.
\end{flushleft}
\end{table*}

The spectra were fitted using the \emph{XSPEC} v12.12.0 software\footnote{https://heasarc.gsfc.nasa.gov/xanadu/xspec} \citep{arnaud96}. We analyzed the \ero\ and \art\ data jointly and the \xmm/EPN data separately. We used the $\chi^2$-statistic to fit the models to the data from the \srg\ telescopes and Cstat for the \xmm/EPN data.

First we tried to fit the source spectrum by a power law with a low-energy cutoff due to photoabsorption in the Galaxy. In the terminology of \emph{XSPEC} we used the following model:
\[
TBabs(zpowerlaw),
\]
where TBabs is the interstellar absorption model by \cite{wilms2000}. Following the example of the authors of previous papers on the X-ray observations of the quasar \pg\ (in particular, \citealt{piconcelli2005}), we fixed the hydrogen column density toward this object at its Galactic value of $N_{\rm H}=5.74\times10^{20}$ cm$^{-2}$ (\citep{elvis1989}.

The results of fitting the the measured spectra by this empirical model are shown in Figs.~\ref{fig:art_ero_sum_spec} and \ref{fig:xmm_sum_spec}. In our modeling we assumed that the slope of the spectrum remained constant in all \srg\ observations, but its normalization (i.e., the X-ray flux) could change. Since in reality the slope could slightly change from observation to observation and given that different observations were carried out at different angles to the optical axis of the telescopes, we added the cross-calibration coefficient between the \ero\ and \art\ telescopes to the model as a free parameter. The same assumption (about the constancy of the spectral slope) was also made with regard to the \xmm\ observations. Thus, we jointly fitted, first, three \srg\ spectra and, second, three \xmm\ spectra. The derived model parameters are given in Table~\ref{tab:fit_pars_art_ero_xmm}. Here and below, we specify the confidence intervals for the parameters at a 68\% level and the upper limits at a 2$\sigma$ level, unless stated otherwise.

Although the power law with absorption, on the whole, describes satisfactorily the shape of the measured spectra, a number of statistically significant additional features are observed: (1) a soft X-ray excess at energies $\sim 0.3$--0.4~keV, (2) an emission excess near 2.7~keV, and (3) a decrease in intensity near 0.5~keV. The first two features manifest themselves in both \srg\ and \xmm\ (less clearly) spectra, while the last one manifests itself in the SRG spectrum taken on November 23, 2019, and at a low confidence level in the October 20, 2019 \srg\ spectrum.

The emission excess at energies below $\sim 1$~keV (in the source rest frame) observed in the spectrum of \pg\ is not unique for this object. Such an additional emission component is detected in the spectra (of good quality) for most Type-1 AGNs, while its nature is actively discussed (see, e.g., \citealt{turner1989,guainazzi2007,boissay2016}). In particular, it can result from the Comptonization of ultraviolet radiation from the accretion disk in the "warm corona" (with a temperature of the order of several million K), in contrast to the main (power-law) component of the hard X-ray emission that is believed to be formed in the "hot corona" ($T\sim 10^9$~K) of the accretion disk.

The energy of 2.7~keV, at which an emission excess is also observed in the spectrum of the quasar \pg, roughly corresponds to the position of the K$\alpha$ line of neutral or weakly ionized iron (6.4~keV) at the redshift of the object: $6.4/(1+z)=2.74$~keV. This suggests that we are dealing with an iron emission line or a set of such lines. As regards the "absorption" feature near 0.5~keV in the \ero\ spectra, it, along with the emission excess at low energies, can be a manifestation of a more complex spectral shape than the simple continuum models that we we tried to apply.

The insufficiently high statistical significance of the detection of the soft emission component in the spectrum of the quasar \pg\ does not allow it to be studied in more detail. Therefore, we excluded the range of energies below 0.7~keV from our subsequent consideration by concentrating on the study of the spectral continuum shape in the energy range from 0.7 to $\sim 13$~keV and the emission feature near the Fe K$\alpha$ line.

\begin{figure*}
    \centering
    \includegraphics[width=0.9\columnwidth]{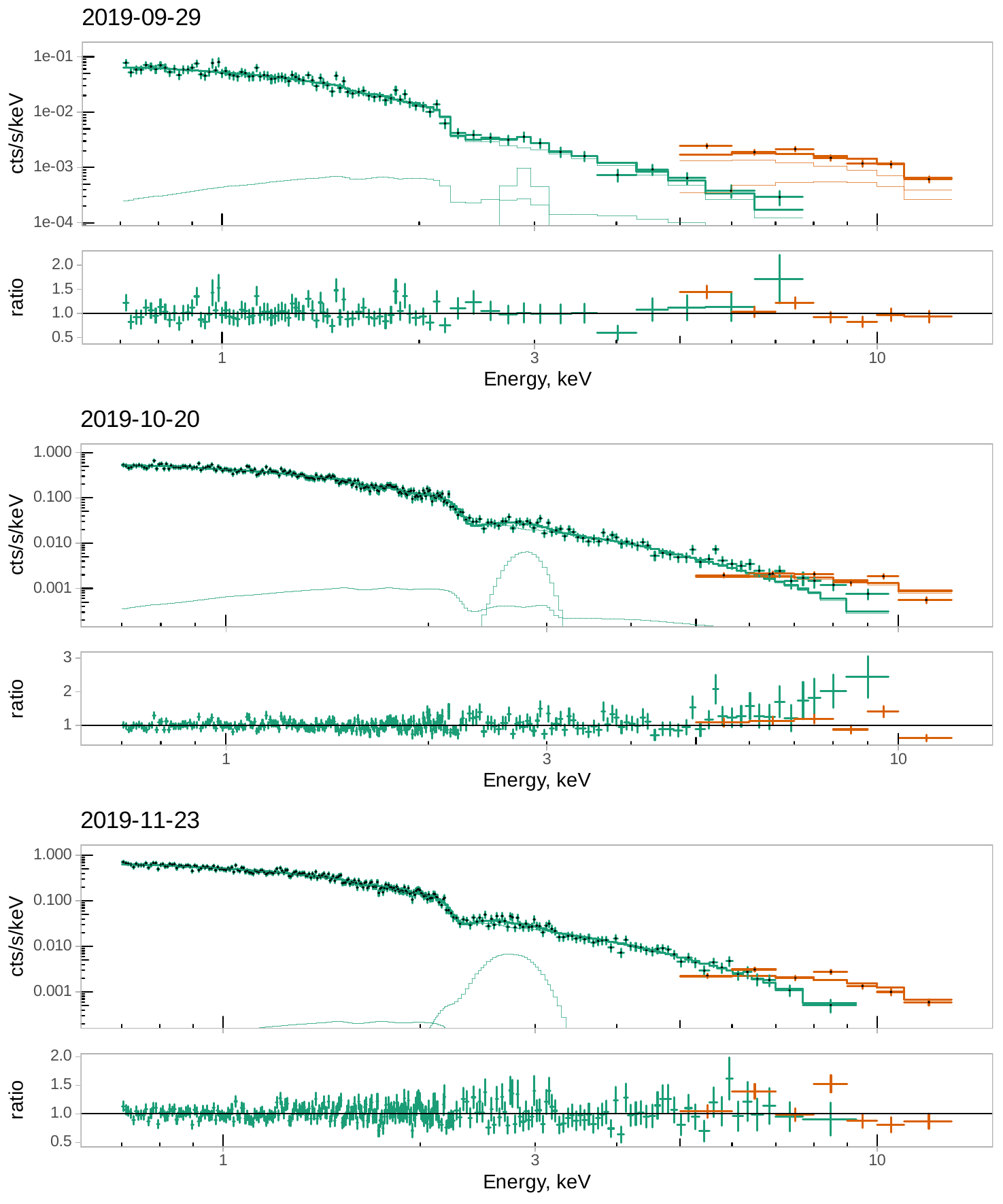}
    \caption{
The fits to the spectra of the quasar \pg\ measured with the \ero\ (green color) and \art\ (orange color) telescopes of the \srg\ observatory at different dates by the \emph{PL+GAUSS+PEXRAV} model with Galactic absorption (see Table~\ref{tab:fit_pars_art_ero}). The solid lines, except for the combined model, indicate its individual components: the power-law continuum (\emph{PL}), the Gaussian line (\emph{GAUSS}), and the reflected component (\emph{PEXRAV}) (if present). The spectra were rebinned for clarity. The data-to-model ratio is shown under each spectrum.
}
    \label{fig:specs_art_ero_by_date}
\end{figure*}

\begin{table*}
\renewcommand{\arraystretch}{1.2}
\caption{
\label{tab:fit_pars_art_ero} 
Results of fitting the measured SRG spectra by different models
}
\centering
\begin{tabular}[t]{llll}
\toprule
Date & 2019-09-29 & 2019-10-20 & 2019-11-23\\
Energy range, keV & 0.7--13 keV & 0.7--12 keV & 0.7--13 keV \\

\midrule
\multicolumn{4}{c}{\emph{PL} model}\\
\midrule
$\Gamma$ & $1.78\pm0.02$ & $1.864\pm0.015$ & $1.850_{-0.013}^{+0.014}$\\
$F_{\rm PL}$ (4--12 keV), $10^{-13}$~erg/s/cm$^2$ & $5.7\pm0.2$ & $5.74\pm0.15$ & $7.05\pm0.16$\\
$\chi^2$ (dof) & $98.8~(91)$ & $316.2~(263)$ & $326~(272)$\\
\midrule
\multicolumn{4}{c}{\emph{PL+GAUSS} model}\\
\midrule
$\Gamma$ & $1.78\pm0.02$ & $1.876\pm0.016$ & $1.866\pm0.014$\\
$F_{\rm PL}$ (4--12 keV), $10^{-13}$~erg/s/cm$^2$ & $5.7\pm0.2$ & $5.60\pm0.16$ & $6.82\pm0.17$\\
$E_{\rm line}$, keV & $6.8_{-0.4}^{+0.2}$ & $6.52_{-0.13}^{+0.11}$ & $6.44\pm0.16$\\
$\sigma_{\rm line}$, keV & $< 0.6$ (68\%) & $0.30_{-0.12}^{+0.20}$ & $0.52_{-0.13}^{+0.12}$\\
$EW$, keV & $0.08\pm0.07$ & $0.11_{-0.04}^{+0.03}$ & $0.14\pm0.04$ \\
$\chi^2$ (dof) & $97~(88)$ & $302.1~(260)$ & $306.7~(269)$\\
\midrule
\multicolumn{4}{c}{\emph{PL+GAUSS+PEXRAV} model}\\
\midrule
$\Gamma$ & $1.95_{-0.08}^{+0.09}$ & $1.90\pm0.03$ & $1.870_{-0.018}^{+0.025}$\\
$F_{\rm PL}$ (4--12 keV), $10^{-13}$~erg/s/cm$^2$ & $4.1_{-0.7}^{+0.6}$ & $5.4\pm0.3$ & $6.8_{-0.3}^{+0.2}$\\
$E_{\rm line}$, keV & $6.8\pm0.3$ & $6.51_{-0.14}^{+0.12}$ & $6.44_{-0.16}^{+0.15}$\\
$\sigma_{\rm line}$, keV & $< 0.8$ (68\%) & $0.32_{-0.13}^{+0.23}$ & $0.51_{-0.11}^{+0.14}$\\
$EW$, keV & $0.12\pm0.08$ & $0.12\pm0.04$ & $0.14_{-0.06}^{+0.04}$\\
$R_{\rm refl}$ & $0.19_{-0.11}^{+0.08}$ & $<0.12$ & $<0.07$  \\ 
$\chi^2$ (dof) & $89.7~(87)$ & $301.1~(259)$ & $306.7~(268)$\\
\bottomrule
\end{tabular}
\begin{flushleft}
$\Gamma$ and $F_{\rm PL}$ are the slope of the power-law component and the Galactic extinction-corrected flux in the observed 4--12~keV energy band in this component; $E_{\rm line}$, $\sigma_{\rm line}$ and $EW$ are the central energy, width, and equivalent width of the Gaussian line; $R_{\rm refl}$ is the reflection coefficient; dof is the number of degrees of freedom.
\end{flushleft}
\end{table*}

\begin{figure*}
    \centering
    \includegraphics[width=0.9\columnwidth]{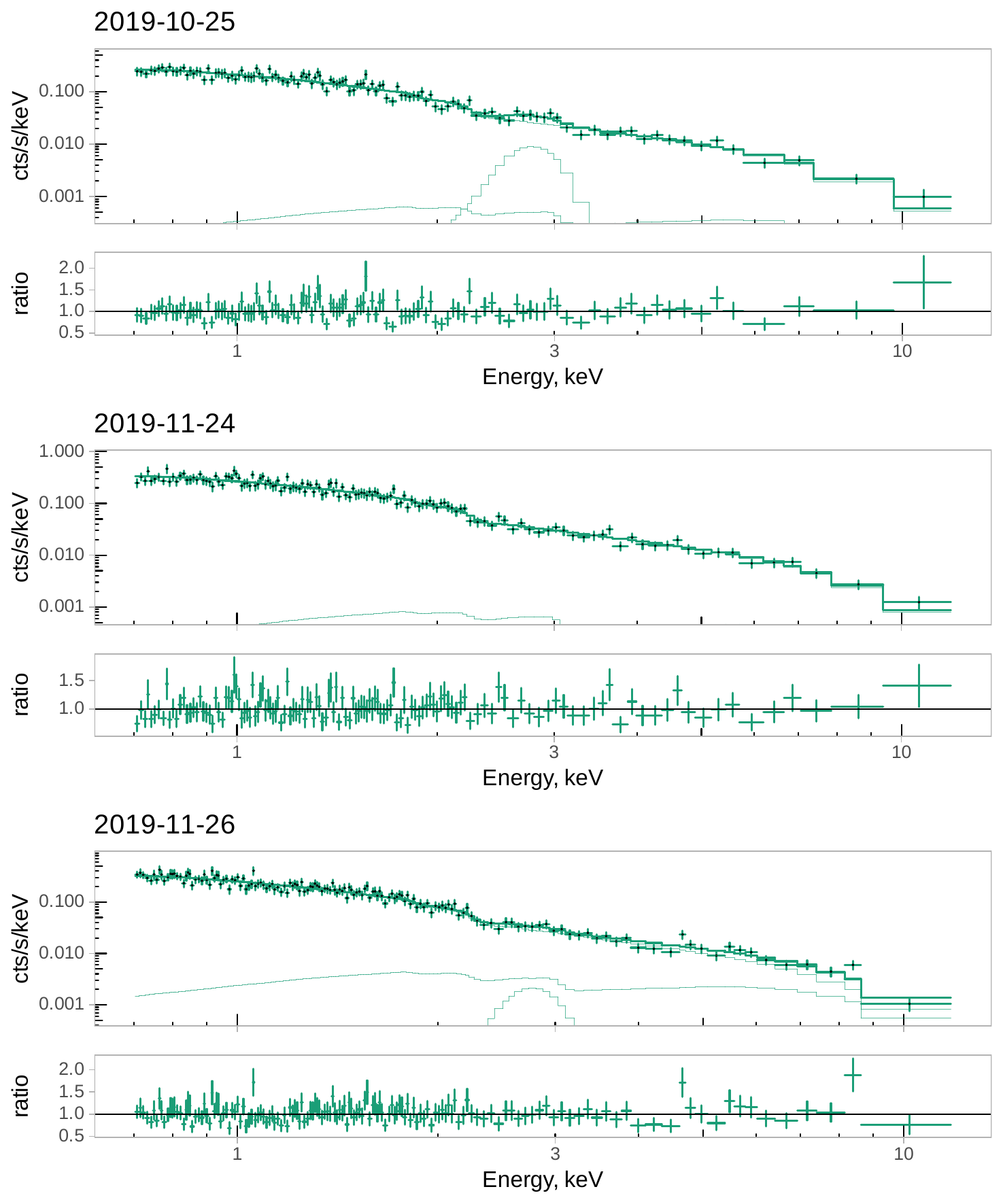}
\caption{
Same as Fig.~\ref{fig:specs_art_ero_by_date} but for the \xmm\ data.
}
    \label{fig:specs_xmm_by_date}
\end{figure*}

\begin{table*}
\renewcommand{\arraystretch}{1.2}
\caption{\label{tab:fit_pars_xmm} 
Results of fitting the measured \xmm\ spectra by different models
}
\centering
\begin{tabular}[t]{llll}
\toprule
Date & 2019-10-25 & 2019-11-24 & 2019-11-26\\
Energy range, keV & 0.7--11.8 keV & 0.7--11.8 keV & 0.7--11.8 keV \\

\midrule
\multicolumn{4}{c}{\emph{PL} model}\\
\midrule
$\Gamma$ & $1.78\pm0.03$ & $1.79\pm0.03$ & $1.78\pm0.03$\\
$F_{\rm PL}$ (4--12 keV), $10^{-13}$~erg/s/cm$^2$ & $6.8\pm0.3$ & $8.4\pm0.4$ & $8.3_{-0.3}^{+0.4}$\\
Cstat (dof) & $562.5~(584)$ & $547.7~(574)$ & $710.9~(747)$\\

\midrule
\multicolumn{4}{c}{\emph{PL+GAUSS} model}\\
\midrule
$\Gamma$ & $1.81\pm0.03$ & $1.79\pm0.03$ & $1.79\pm0.03$\\
$F_{\rm PL}$ (4--12 keV), $10^{-13}$~erg/s/cm$^2$ & $6.4_{-0.3}^{+0.4}$ & $8.4\pm0.4$ & $8.2\pm0.4$\\
$E_{\rm line}$, keV & $6.5$ (fixed) & $6.5$ (fixed) & $6.5$ (fixed)\\
$\sigma_{\rm line}$, keV & $0.50$ (fixed) & $0.50$ (fixed) & $0.50$ (fixed)\\
$EW$, keV & $0.20\pm0.06$ & $<0.11$ & $<0.13$ \\
Cstat (dof) & $551.7~(583)$ & $547.7~(573)$ & $710.6~(746)$\\

\midrule
\multicolumn{4}{c}{\emph{PL+GAUSS+PEXRAV} model}\\
\midrule
$\Gamma$ & $1.83_{-0.05}^{+0.06}$ & $1.81_{-0.04}^{+0.05}$ & $1.90\pm0.06$\\
$F_{\rm PL}$ (4--12 keV), $10^{-13}$~erg/s/cm$^2$ & $6.1_{-0.7}^{+0.6}$ & $8.0_{-0.8}^{+0.7}$ & $6.6\pm0.7$\\
$E_{\rm line}$, keV & $6.50$ (fixed) & $6.50$ (fixed) & $6.50$ (fixed)\\
$\sigma_{\rm line}$, keV & $0.50$ (fixed) & $0.50$ (fixed) & $0.50$ (fixed)\\
$EW$, keV & $0.20_{-0.06}^{+0.05}$ & $<0.08$  & $<0.17$  \\
$R_{\rm refl}$ & $<0.3$ & $<0.2$ & $0.22_{-0.11}^{+0.09}$\\
Cstat (dof) & $551.5~(582)$ & $547.4~(572)$ & $704.4~(745)$\\
\bottomrule
\end{tabular}
\end{table*}

The subsequent analysis was performed separately for each \srg\ and \xmm\ spectrum, i.e., we assumed that not only the normalization but also the spectral shape could change from observation to observation. The cross-calibration constant between the \ero\ and \art\ data was no longer used.

We used three models. First, we again ap- plied the power-law model with Galactic absorption ($TBabs(zpowerlaw)$ in the terminology of \emph{XSPEC}) below called the \emph{PL} model. Then, to describe the emission feature near 6.4~keV (in the quasar rest frame), we added a line with a Gaussian profile to this model: $TBabs(zpowerlaw + zgauss)$ (hereafter \emph{PL+GAUSS}).

If the observed iron emission line is associated with the reflection of hard X-ray coronal radiation from the accretion disk or the dusty torus, then it is natural to also expect an additional emission in spectral continuum (at energies above $\sim 10$~keV in the quasar rest frame). 
Therefore, we considered one more model, where the component associated with the reflection of a power-law continuum from a cold disk \citep{magdziarz1995} was added to the two previous components, i.e., we used the $TBabs(zpowerlaw + zgauss + pexrav)$ or \emph{PL+GAUSS+PEXRAV} model. The relative normalization of the reflected PEXRAV component with respect to the power-law PL component was described by the parameter $rel_{\rm refl}$ that for this purpose was specified to be negative (see the description of PEXRAV in \emph{XSPEC}); below we will use the notation $R_{\rm refl}\equiv -rel_{\rm refl}$. The spectral slope of the incident radiation in the \emph{PEXRAV} component was tied to the slope of the \emph{PL} component. The abundance of heavy elements was fixed at the solar one, according to \cite{anders1989}. The cosine of the inclination angle was fixed at 0.5. 
The high-energy cutoff was not considered. 
Because of the low detection significance of the iron emission line in the \xmm\ spectra, when fitting these spectra by the \emph{PL+GAUSS} and \emph{PL+GAUSS+PEXRAV} models, the central energy and width of the line were fixed at the weighted mean values obtained when fitting the \srg\ spectra by the \emph{PL+GAUSS+PEXRAV} model.

Figures~\ref{fig:specs_art_ero_by_date} and \ref{fig:specs_xmm_by_date} show how the \srg\ and \xmm\ spectra are fitted by the three-component \emph{PL+GAUSS+PEXRAV} model. Tables \ref{tab:fit_pars_art_ero} and \ref{tab:fit_pars_xmm} present the parameters of the fits to the spectra by the \emph{PL}, \emph{PL+GAUSS} and \emph{PL+GAUSS+PEXRAV}.

Adding the iron emission line to the power-law model leads to a statistically significant improvement in the quality of the fit to the \srg\ spectra taken on October 20 and November 23, 2019 (recall that the spectrum constructed from the September 29 observation is characterized by considerably poorer statistics, because only one of the \ero\ modules operated in this observation) and the \xmm\ spectrum measured on October 25, 2019. The line width from the \srg/\ero\ data turns out to be nonzero, while the position of the line center is limited in the range from $\sim 6.3$ to $\sim 6.6$~keV for the October 20 and November 23, 2019 spectra, i.e., it is compatible with an energy of 6.4~keV.

The reflected continuum component (\emph{PEXRAV}) is not detected at a statistically significant level in any of the spectra obtained, except for the latest \xmm\ observation (on November 26, 2019), where there is weak evidence (at a level of $\sim 2$ standard deviations) for its presence.

\begin{figure*}
    \centering
    \includegraphics[width=.9\columnwidth]{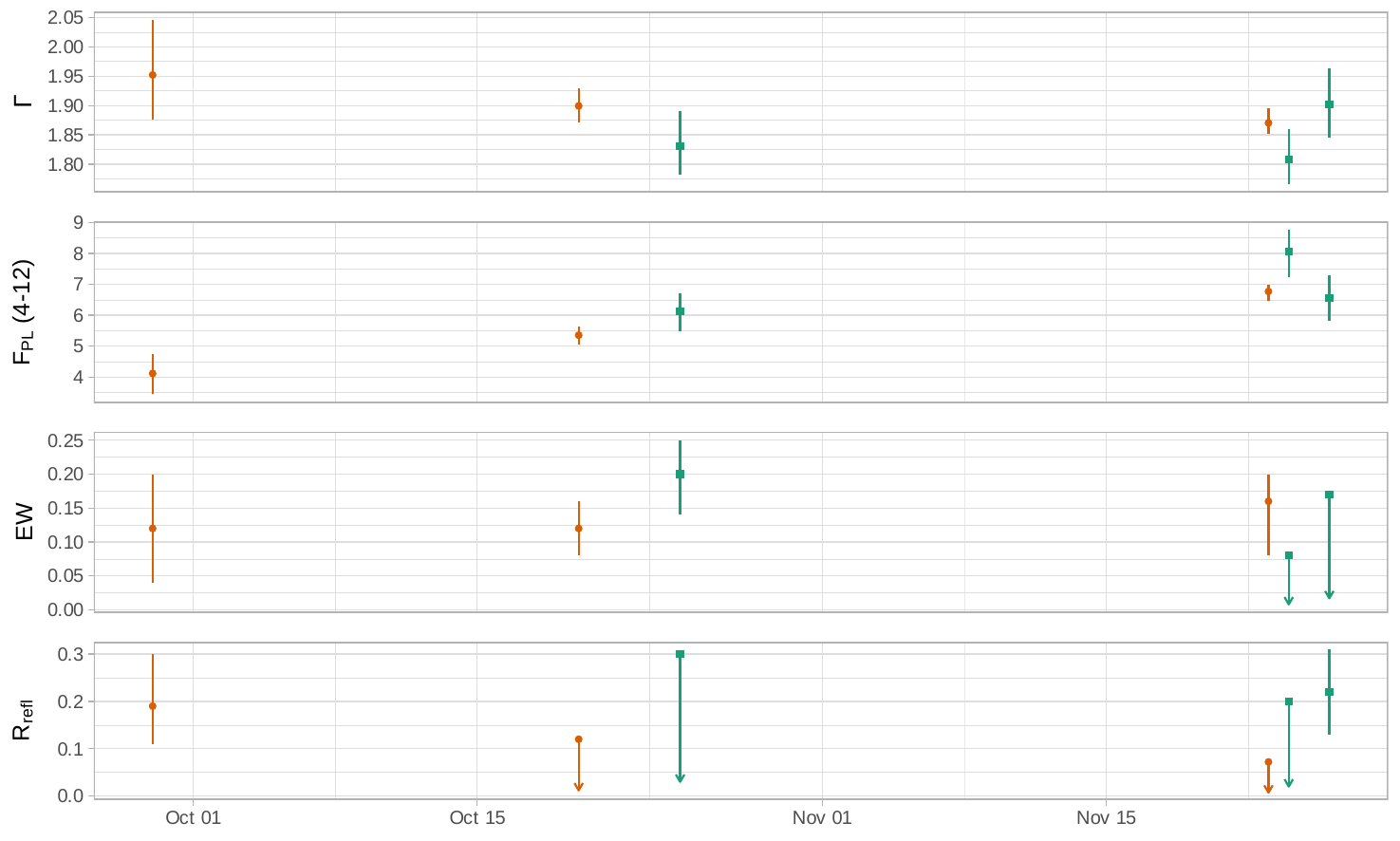}
    \caption{
    Variations in the parameters of the \emph{PL+GAUSS+PEXRAV} spectral model during the \srg\ (orange color) and \xmm\ (green color) observations in the fall of 2019. The slope of the power-law component, the flux in the power-law component (in units of $10^{-13}$~erg/s/cm$^2$), the equivalent width of the iron line, and the relative normalization of the reflected component (see Tables \ref{tab:fit_pars_art_ero} and \ref{tab:fit_pars_xmm}) are shown (from top to bottom).
    }
    \label{fig:lc_fit_params}
\end{figure*}

Figure~\ref{fig:lc_fit_params} shows how the parameters of the \emph{PL+GAUSS+PEXRAV} model changed from observation to observation. The slope of the spectrum (power-law component) and the equivalent width of the iron line do not show any statistically significant evolution, remaining at $\sim 1.9$ and $\sim 120$~eV, respectively. At the same time, there is a statistically significant flux increase in the power-law spectral component.

\section{LONG-TERM EVOLUTION}

\begin{figure*}
    \centering
    \includegraphics[width=0.9\columnwidth]{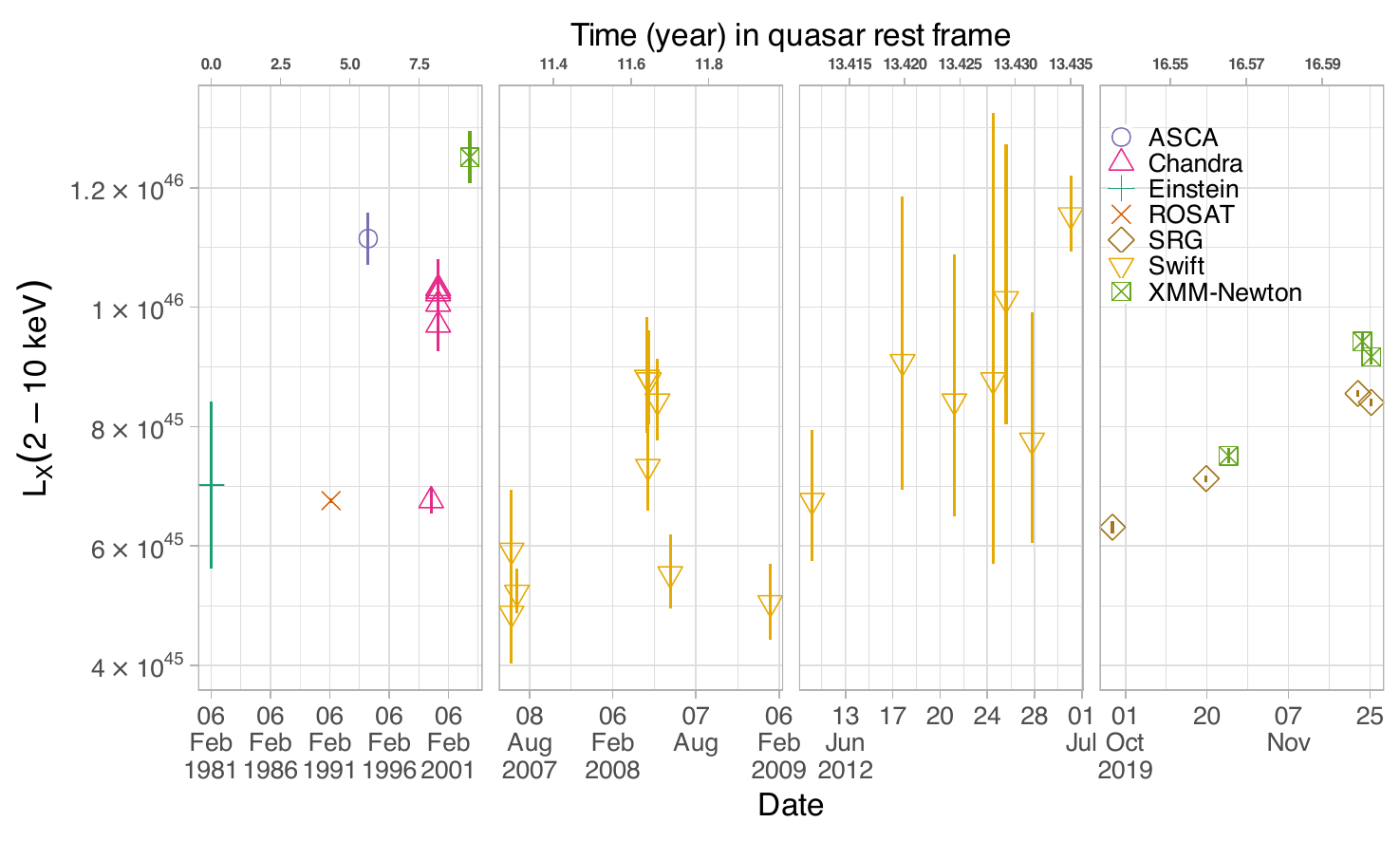}
    \caption{
    Long-term evolution of the X-ray luminosity of the quasar \pg\. The luminosity is given in the 2--10~keV energy band in the source rest frame. The 1$\sigma$ statistical measurement errors are shown. The date of observations is along the lower axis; the time since the first observation in the source rest frame is along the upper axis.
    }
    \label{fig:lc_long_term}
\end{figure*}

As has already been noted in the Introduction, the quasar \pg\ has been repeatedly observed by different X-ray observatories since 1981. This allows the evolution of its activity in X-rays to be traced over approximately 17 years in the source rest frame. Brief information about all X-ray observations of \pg\, namely the Einstein, ROSAT, ASCA, Chandra, XMM-Newton, Swift, and \srg\ observations, is collected in Table~\ref{tab:xray_obs}.

\cite{shemmer2014} estimated the X-ray fluxes in the observed 0.2--10~keV energy band corrected for Galactic absorption for all observations until 2012 inclusive. The quasar spectrum was assumed to be fitted by a power law with a photon index of $\Gamma = 2$. Only the \srg\ and \xmm\ observations carried out in the fall of 2019 are the new ones compared to the paper by \cite{shemmer2014}. We used the X-ray fluxes from \cite{shemmer2014} to estimate the quasar luminosity in the 2--10~keV energy band (in the source rest frame) for all archival observations. Using the same spectral model (a power law with $\Gamma=2$ and Galactic absorption) and the data being discussed in this paper, we calculated the luminosities for the \srg\ and \xmm\ observations in 2019. Our luminosity estimates and their statistical errors are given in Table~\ref{tab:xray_obs}. Note that these errors turned out to be considerably smaller ($\lesssim 2$\%) than the corresponding errors in the flux of the power-law component obtained by us previously when modeling the \srg\ and \xmm\ spectra (cf. the results in Tables \ref{tab:fit_pars_art_ero} and \ref{tab:fit_pars_xmm}). This is because in that case the spectral slope was a free parameter and a number of free parameters were also used simultaneously.

It should be noted that since different measurements were obtained from the data of different instruments and in originally different energy bands, this must lead to an additional systematic error in the luminosity estimate. This uncertainty probably does not exceed 10--20\% \citep{shemmer2014} for the data obtained after 2000, given the careful cross-calibration of the modern X-ray observatories and that all these data were obtained in approximately identical energy bands. However, the systematic error can be more significant for earlier observations, in particular, the Einstein and ROSAT ones, especially when the soft X-ray band (0.3--3.3 and 0.5--2.0~keV, respectively) in which these measurements were carried out is taken into account.

Figure~\ref{fig:lc_long_term} shows how the luminosity of the quasar \pg\ evolved in 1981--2019. Although the light curve is characterized by a large sampling interval, it can be said with confidence that the X-ray emission is variable on time scales from several days to several years (in the source rest frame). The ratio of the maximum ($\sim 1.25\times 10^{46}$~erg/s in November 2002) to minimum ($\sim 5\times 10^{45}$~erg/s in January 2009) X-ray luminosity in the entire history of observations of the quasar is $\sim 2.5$. At the same time, $
\sigma^2_{\rm rms}=0.030\pm0.016$ (from Eq. \ref{eq:sigma}), i.e., the rms
characteristic amplitude of the luminosity variations was $\sim 16$\%. In the fall of 2019, during the \srg\ and \xmm\ observations, the object was in an "average" state for itself, when its luminosity varied in the range from $6\times 10^{45}$ to $9\times 10^{45}$~erg/s and rose almost monotonically.

\section{DISCUSSION AND CONCLUSIONS}

Although the quasar \pg\ has already been repeatedly studied in X-rays previously, its new \srg\ and \xmm\ observations have allowed us to measure more accurately its energy spectrum in the wide range $\sim 1$--30~keV (in the quasar rest frame). One of the most interesting results was the detection of a broad ($\sim 1$~keV at half maximum) iron emission line in the spectrum with an equivalent width $\sim 120$~eV. The weighted mean position of the line centroid from different observations is consistent with the energy of 6.4~keV corresponding to the $K_\alpha$ transition in the neutral iron atom (and is inconsistent with the energy of 6.7~keV of the helium-like iron triplet in the case of hot rarefied plasma radiation). The line broadening is statistically significant.

It is natural to assume that this emission feature is associated with the reflection of hard X-ray coronal radiation from the accretion disk and, possibly, the dusty torus. In this case, the reflected component must also emerge in the spectral continuum. The search for this component based on the \srg\ and \xmm\ data did not lead to a significant detection. The reflection coefficient in the \emph{PEXRAV} model is $R_{\rm refl}\lesssim 0.3$. This limit and the measured equivalent width of the iron emission line are consistent with the scenario for the reflection of coronal radiation from an optically thick, cold disk (see, e.g., \citealt{george1991}). In this case, the significant line broadening can be associated with the Doppler broadening in the accretion disk (this requires radial velocities $v_{\rm r}/c\sim 0.06$), as is often discussed in the context of X-ray binaries and AGNs (for a review, see, e.g., \citealt{miller2007}), and with the reflection from a strongly ionized gas in the inner disk (in this case, a set of K$\alpha$ lines of different iron ions arises; see, e.g., \citealt{nayakshin2000,ross2005}).

As regards the main (power-law) component of the spectrum, the values of its slope measured in different \srg\ and \xmm\ observations agree with each other within the error limits and lie in the range from 1.8 to 2.0. Such values are typical for Seyfert galaxies and moderate-luminosity quasars. Thus, for the quasar \pg\ we see no confirmation of the tendency for the X-ray continuum to steepen significantly with increasing luminosity and/or Eddington ratio that was pointed out for AGNs by a number of authors \citep{shemmer2008,brightman2013} but was called into question in other papers (see, e.g., \citealt{trakhtenbrot2017}).

Apart from the spectral properties, we investigated the X-ray variability of the quasar \pg. On time scales of the order of several hours (here and below, all time intervals are given in the quasar rest frame) its X-ray luminosity exhibits no statistically significant variability. However, the luminosity changed noticeably from observation to observation in the fall of 2019, having increased approximately by 50\% in $\sim 25$~days. A comparison of these new \srg\ and \xmm\ measurements with the previous measurements of other X-ray observations showed that in the entire 17-year history of observations of the quasar \pg\ its X-ray luminosity varied by no more than a factor of 2.5, while the variations on time scales of several weeks and several years are comparable in amplitude.

It is interesting to consider these results in the general context of X-ray AGN variability. Recently, based on a representative sample of X-ray bright quasars from SDSS, \srg/\ero\ all-sky survey data, and archival \xmm\ data, \citealt{prokhorenko2021} established that the characteristic X-ray variability amplitude for quasars increases slowly with time but decreases with luminosity. For quasars with a luminosity $\sim 10^{46}$~erg/s the characteristic variability amplitude (the ratio of a random pair of fluxes) on time scales of 10--20 years is $\sim 1.4$ with a significant dispersion from object to object. Thus, the quasar \pg\ does not look remarkable as regards the X-ray variability among quasars with a comparable luminosity.

\section{Acknowledgements}
In this study we used data from the \art\ and \ero\ telescopes onboard the \srg\ observatory. The \srg\ observatory was designed by the Lavochkin Association (enters into the Roskosmos State Corporation) with the participation of the Deutsches Zentrum fur Luft- und Raumfahrt (DLR) within the framework of the Russian Federal Space Program on the order of the Russian Academy of Sciences. The \ero\ X-ray telescope was built by a consortium of German Institutes led by MPE, and supported by DLR. The \art\ team thanks the Roskosmos State Corporation, the Russian Academy of Sciences, and the Rosatom State Corporation for supporting the design and production of the \art\ telescope and the Lavochkin Association and partners for the production and work with the spacecraft and the Navigator platform. The \ero\ data used in this work were processed with the eSASS software developed by the German \ero\ consortium and the proprietary data reduction and analysis software developed by the Russian \ero\ Consortium.

\section{FUNDING}

This study was supported by RSF grants nos. 21-12-00343 and 19-12-00396 with regard to the eROSITA and ART-XC data processing, respectively.

\bibliographystyle{mnras}
\bibliography{main}

\section{APPENDIX}

\begin{table*}
\caption{
\label{tab:xray_obs}
History of X-ray observations of the quasar \pg
}
\centering
\renewcommand{\arraystretch}{1.2}
\begin{tabular}[t]{lcccll}
\toprule
Date & Observatory & OBSID & Exposure time, ks & References & $L_{\rm X}$, $10^{45}$~erg/s\\
\midrule
1981-02-06 & Einstein & 5351 & 1.83 & 1, 4, 9 & $7.0_{-1.4}^{+1.4}$\\
1991-03-15 & ROSAT & 700246 & 9.01 & 9 & $6.76_{-0.04}^{+0.04}$\\
1994-05-02 & ASCA & 71036000 & 47.7 & 2, 3, 9 & $11.1_{-0.4}^{+0.4}$\\
1999-08-21 & Chandra & 1269 & 10.83 & 7, 8, 9 & $6.8_{-0.2}^{+0.2}$\\
2000-03-23 & Chandra & 47 & 5.39 & 8, 9 & $10.2_{-0.4}^{+0.4}$\\
2000-03-23 & Chandra & 62 & 4.85 & 8, 9 & $10.3_{-0.4}^{+0.4}$\\
2000-03-24 & Chandra & 69 & 4.86 & 8, 9 & $10.1_{-0.4}^{+0.4}$\\
2000-03-24 & Chandra & 70 & 4.86 & 8, 9 & $9.7_{-0.4}^{+0.4}$\\
2000-03-24 & Chandra & 71 & 4.41 & 8, 9 & $10.3_{-0.5}^{+0.5}$\\
2002-11-22 & XMM-Newton & 143150101 & 13.7 & 4, 6, 5, 9 & $12.5_{-0.4}^{+0.4}$\\
2007-06-29 & Swift & 36672001 & 1.32 & 9 & $4.9_{-0.8}^{+1.0}$\\
2007-06-29 & Swift & 36673001 & 1.48 & 9 & $5.9_{-0.9}^{+1.0}$\\
2007-07-11 & Swift & 36672002 & 7.34 & 9 & $5.2_{-0.4}^{+0.4}$\\
2008-04-22 & Swift & 36671002 & 2.09 & 9 & $8.8_{-0.9}^{+1.0}$\\
2008-04-24 & Swift & 36673002 & 2.56 & 9 & $7.3_{-0.7}^{+0.8}$\\
2008-04-26 & Swift & 36671003 & 3 & 9 & $8.8_{-0.7}^{+0.8}$\\
2008-05-15 & Swift & 90030001 & 3.72 & 9 & $8.4_{-0.7}^{+0.7}$\\
2008-06-12 & Swift & 90030002 & 3.25 & 9 & $5.5_{-0.6}^{+0.7}$\\
2009-01-18 & Swift & 90030003 & 2.75 & 9 & $5.0_{-0.6}^{+0.7}$\\
2012-06-11 & Swift & 91438001 & 1.26 & 9 & $6.8_{-1.0}^{+1.2}$\\
2012-06-18 & Swift & 91438002 & 0.37 & 9 & $9_{-2}^{+3}$\\
2012-06-22 & Swift & 91438004 & 0.42 & 9 & $8.4_{-1.9}^{+2.5}$\\
2012-06-25 & Swift & 91438005 & 0.17 & 9 & $9_{-3}^{+4}$\\
2012-06-26 & Swift & 91438006 & 0.36 & 9 & $10_{-2}^{+3}$\\
2012-06-28 & Swift & 91438007 & 0.5 & 9 & $7.8_{-1.7}^{+2.2}$\\
2012-07-01 & Swift & 91438008 & 5.88 & 9 & $11.5_{-0.6}^{+0.7}$\\
2019-09-29 & \srg &  & 48.3, 55.0 & 10 & $6.31_{-0.11}^{+0.11}$\\
2019-10-20 & \srg &  & 37.7, 39.0 & 10 & $7.13_{-0.05}^{+0.05}$\\
2019-10-25 & XMM-Newton & 852980501 & 14.4 & 10 & $7.51_{-0.13}^{+0.13}$\\
2019-11-23 & \srg &  & 37.0, 38.3 & 10 & $8.55_{-0.06}^{+0.06}$\\
2019-11-24 & XMM-Newton & 852980301 & 12.7 & 10 & $9.43_{-0.15}^{+0.15}$\\
2019-11-26 & XMM-Newton & 852980401 & 17.3 & 10 & $9.16_{-0.13}^{+0.13}$\\
2019-11-26 & \srg &  & 39.8 & 10 & $8.40_{-0.07}^{+0.07}$\\
\bottomrule
\end{tabular}
\begin{flushleft}
    References: (1) \cite{tananbaum1986}, (2) \cite{nandra1995}, (3) \cite{george2000}, (4) \cite{page2004}, (5) \cite{jimenez2005}, (6) \cite{piconcelli2005}, (7) \cite{harocorzo2007}, (8) \cite{park2008}, (9) \cite{shemmer2014}, (10) this paper. $L_{\rm X}$ is the Galactic absorption-corrected luminosity in the 2--10~keV energy band in the quasar rest frame.
\end{flushleft}
\end{table*}

\end{document}